\documentclass[reprint,twocolumn,aps,prl,superscriptaddress]{revtex4-2}
\usepackage{mhchem}
\usepackage{fancyhdr}
\usepackage{bigdelim}
\usepackage{float}
\usepackage[english]{babel}
\usepackage[utf8]{inputenc}
\usepackage{fancyhdr}
\usepackage{sidecap}
\usepackage{multirow}

\usepackage{graphicx}
\usepackage{tabularx}
\usepackage{braket}
\usepackage{tikz}
\usepackage{color}
\usepackage{amsmath}
\usepackage[colorlinks=True, linkcolor=blue, filecolor=magenta, urlcolor=blue,citecolor=blue]{hyperref}
\usepackage[nameinlink,capitalise]{cleveref}

\begin{document}
\title{Field-driven phase transition crossover in kagome antiferromagnet Mn$_{3-x}$Co$_x$Sn}

\author{Shubhankar Paul}
\email{shubhp@iitk.ac.in}
\affiliation{Department of Physics, Indian Institute of Technology Kanpur, Kanpur 208016, India}
\author{Chanchal Sow}
\email{chanchal@iitk.ac.in}
\affiliation{Department of Physics, Indian Institute of Technology Kanpur, Kanpur 208016, India}
	
\date{\today}
\begin{abstract}
Mn$_3$Sn exhibits a complex magnetic phase diagram with multiple symmetry-breaking phenomena.
Minute electron doping at the Mn site breaks the kagome lattice symmetry in Mn$_3$Sn. 
In this work, we exploited this to tune the order of the magnetic phase transition using magnetocaloric effect in Mn$_{3-x}$Co$_x$Sn single crystals.
Pristine Mn$_3$Sn shows metallic behaviour with a spin-reorientation transition at 278 K, which shifts to 175 K upon Co doping.
Additionally, Co doping also induces a weak-ferromagnetic transition near $\approx 210$ K.
The effect of Co doping results inverse to conventional magnetocaloric effect with enhanced magnetic entropy change.
Further, the scaled entropy change reveals a field-driven magnetic phase transition crossover from first order to second order in Mn$_{2.75}$Co$_{0.25}$Sn crystal. Based on experimental data, a new phase diagram is constructed, depicting the field evolution of various magnetic orders.

\end{abstract}

\maketitle

\section{Introduction}
Phase transitions in antiferromagnets (AFMs) often involve strong coupling among structural, magnetic, and electronic degrees of freedom, leading to complex phase diagrams \cite{kawamura1984phase,vopson2015fundamentals,singh2024higher}.
In such systems, evaluating the magnetic phase diagram and order of phase transition requires bulk thermodynamic measurements on high-quality single crystals.
In this regard, magnetocaloric effect (MCE) can be a suitable probe to determine the order of the phase transition \cite{tegus2002transition,franco2006field,dong2008phenomenological,franco2010scaling,biswas2013universality,Vopson04072015,law2018quantitative,zhou2022unified}.
The conventional magnetocaloric effect (CMCE) is characterised by the positive value of magnetic entropy change [$-\Delta S_\mathrm{M} (T)$] \cite{krenke2005inverse,biswas2013universality,bonilla2010universal,law2018quantitative}.
\mbox{V. Franco} $et$ $al.$\cite{franco2010scaling,franco2006field,dong2008phenomenological} proposed a scaling approach based on the collapse of many field-dependent $-\Delta S_\mathrm{M} (T)$ curves to a universal master curve in case of second order phase transition (SOPT) \cite{franco2010scaling,franco2006field}. 
The universality of this scaling approach has been widely confirmed in various magnetic systems \cite{franco2010scaling,franco2006field,bonilla2010universal}.
The breakdown of the universal curve is expected for the first-order phase transition (FOPT) \cite{dwari2024evidence,pandey2017large}.
SOPT typically exhibits a modest distribution of $-\Delta S_\mathrm{M} (T)$ over a wide temperature span, whereas a first-order phase transition (FOPT) is distinguished by a sharp peak confined to a narrow temperature window \cite{brown1976magnetic,pandey2017large,mazet2006mn3sn2}.
A positive temperature derivative of magnetisation leads to the inverse magnetocaloric effect (IMCE), characterised by \mbox{$-\Delta S_\mathrm{M} (T)<0$}. 
Such systems often undergo FOPT due to modifications of the magnetic sublattices between AFM to ferrimagnetic \cite{tegus2002magnetic,zhang2004giant}, AFM to FM \cite{nikitin1990magnetocaloric,tohei2003negative,hu2008giant}, and collinear to noncollinear AFM phases \cite{tegus2002magnetic,gottschilch2012study, biniskos2018spin}. 
IMCE in AFMs is commonly related to the temperature- and field-dependent magnetostructural effects \cite{tegus2002magnetic,gottschilch2012study, biniskos2018spin,pandey2017large,dwari2024evidence}.
Mn$_3$Sn is one among such candidates that possess a rich phase diagram and has immense potential in AFM spintronics \cite{higo2018large,kuroda2017evidence,ikhlas2017large,kimata2019magnetic,li2019chiral,cheng2019terahertz,wang2022noncollinear,jungwirth2016antiferromagnetic}.
\begin{figure}[b]
		\begin{center}
			\includegraphics[scale=0.38]{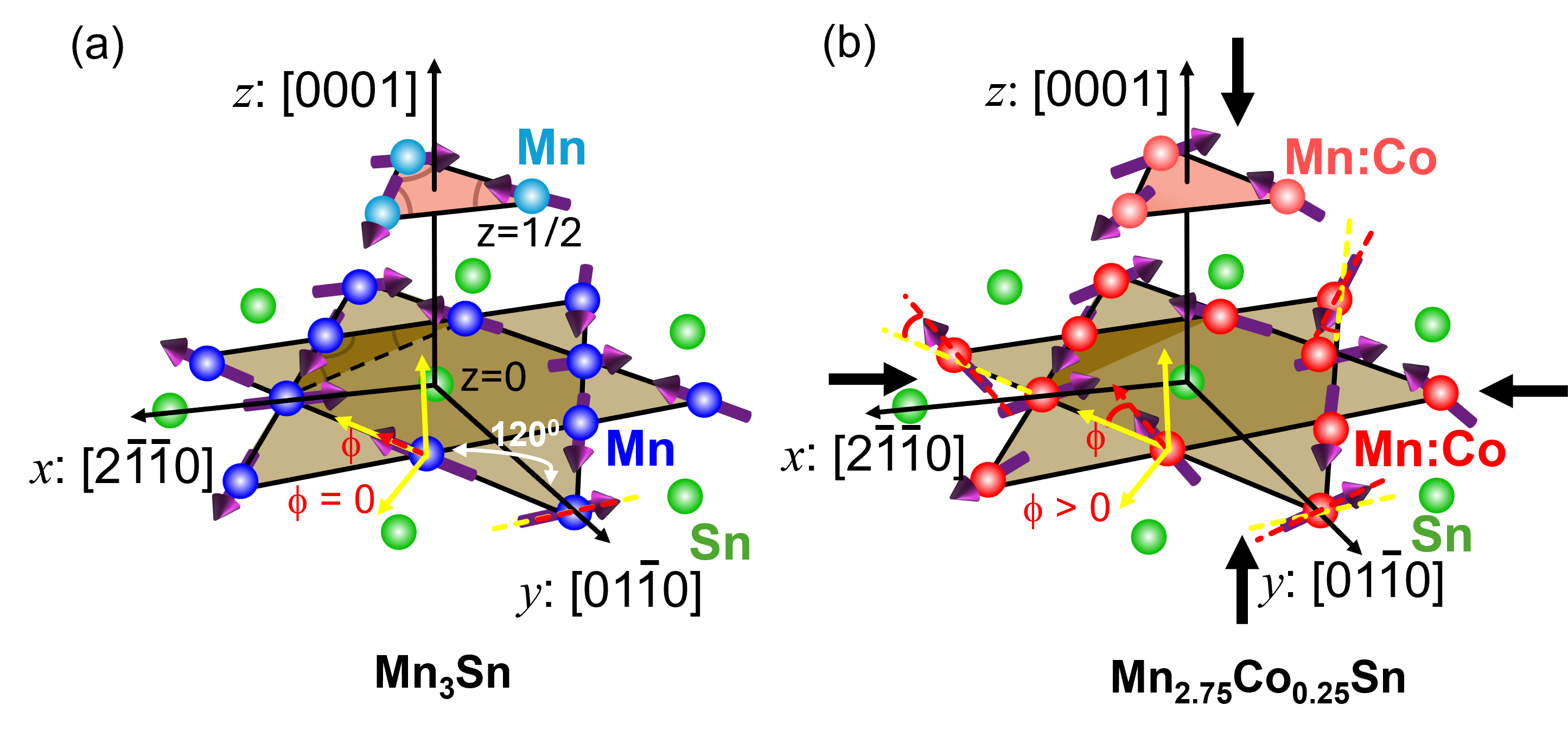}
			\caption{Schematic diagram of the crystal structure and possible magnetic configuration of (a) Mn$_3$Sn and \mbox{(b) Mn$_{2.75}$Co$_{0.25}$Sn}. The Mn magnetic moments confined to the kagome plane (hexagonal $ab$) adopt an inverse triangular spin arrangement (highlighted in $old\ gold$ color). Upon Co doping, the magnetic moments become slightly canted toward the $c$-axis by an angle  $\phi~>~0$.}
			\label{fig0}
		\end{center}
	\end{figure}

Mn$_3$Sn has a hexagonal crystal structure with ABAB-stacking of the (0001) planes.
At room temperature, Mn moments form a noncollinear triangular-spin (NTS) structure \cite{sung2018magnetic,nakatsuji2015large,tomiyoshi1982polarized,chen2021anomalous,tomiyoshi1982polarized,tomiyoshi1982magnetic,OHMORI1987249,radhakrishna1991neutron,brown1990determination} as shown in \cref{fig0}\textcolor{blue}{a}, which transforms into a noncollinear spin-spiral (NSS) structure \cite{tomiyoshi1982polarized,tomiyoshi1982magnetic,OHMORI1987249,radhakrishna1991neutron,brown1990determination,chen2021anomalous}.
Neutron diffraction studies further reveal weak ferromagnetism in the NTS phase, arising from the distortion of the equilateral Mn triangle in the kagome lattice and Mn-rich off-stoichiometry \cite{tomiyoshi1982polarized,tomiyoshi1982magnetic,OHMORI1987249,radhakrishna1991neutron}.
Mn$_3$Sn exhibits a temperature-dependent change of its magnetic structure that can be altered by electron doping, strain/pressure \cite{hanggai2025complex,low2022tuning,low2024topological, TOMIYOSHI19861001}.
In particular, doping with magnetic elements at the Mn sites breaks the kagome lattice symmetry, leading to changes in the Dzyaloshinskii–Moriya interaction (DMI), exchange coupling, and magnetic anisotropy.
Thus, electron doping can be employed to control the magnetic structure, which gives rise to the emergence of competing phases \cite{hanggai2025complex,low2022tuning,low2024topological, singh2024higher}.
Recent work on  Mn$_{2.5}$Fe$_{0.6}$Sn$_{0.9}$ and Mn$_{2.5}$Cr$_{0.6}$Sn$_{0.9}$ support such scenarios \cite{liu2021evolution,hanggai2025complex,singh2024higher,liu2020exchange}.
It is important to note that any magnetic dopant with high concentrations destabilises the underlying AFM structure of the parent system \cite{yano2024tuning,liu2020exchange,hanggai2025complex,liu2020exchange,singh2024higher}.

Here, we study the effect of Co doping (8\%), preserving the kagome AFM structure (\cref{fig0}\textcolor{blue}{b}).
We find a field-driven change in the order of phase transitions using MCE in Mn$_{3-x}$Co$_{x}$Sn ($x$ = 0, 0.25) single crystals with an additional weak ferromagnetic-like (WFM) transition. 
Based on the experimental results, we propose a new magnetic and MCE phase diagram.

\begin{figure*}
		\begin{center}
			\includegraphics[scale=0.68]{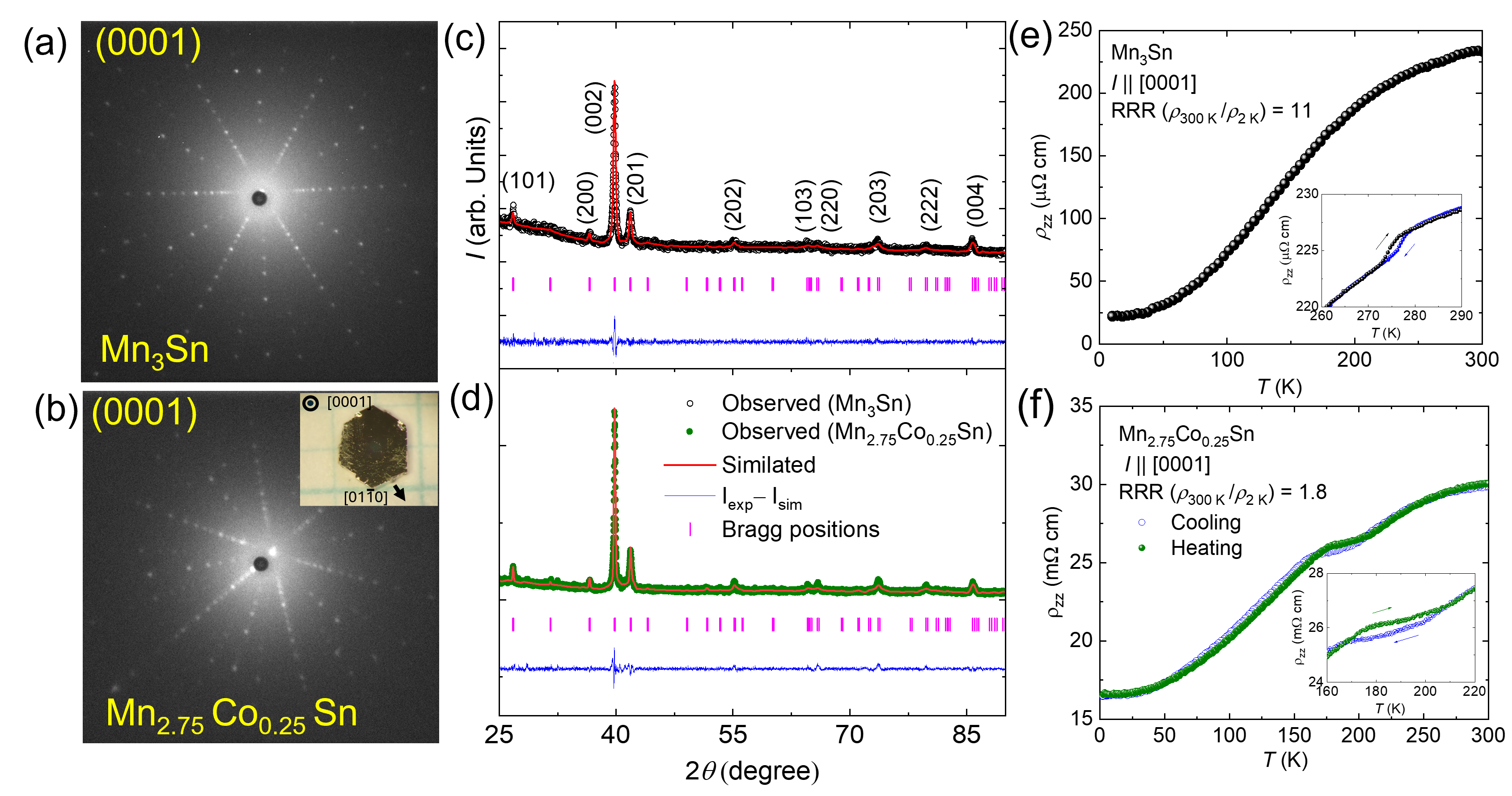}
			\caption{Laue diffraction pattern of (a) Mn$_3$Sn and (b) Mn$_{2.75}\mathrm{Co}_{0.25}$Sn showing (0001) plane. The inset of (b) shows the optical image of a typical Mn$_{2.75}\mathrm{Co}_{0.25}$Sn single crystal. (c-d) Powder X-ray diffraction patterns on crushed single crystals of both systems. (e-f) The temperature-dependent zero-field longitudinal resistivity 
        ($\rho_\mathrm{zz}$) measured with $I\ \|$ [0001] for both crystals. The insets show the zoomed-in view of the $\rho-T$ plot across SRT.}
			\label{fig1}
		\end{center}
	\end{figure*}
\section{Experimental}
Single crystals of Mn$_{3-x}$Co$_{x}$Sn ($x$ = 0, 0.25) were grown using the self-flux growth technique \cite{sung2018magnetic,paul2026CrSb}. High-purity Mn (99.99\%), Co (99.999\%), and Sn (99.99\%) pieces were mixed in a Mn:Co:Sn atomic ratio of (7-$x$):$x$:3 and placed into an alumina crucible. The crucible was then sealed inside a quartz tube under vacuum (10$^{-5}$ mbar). At first, the tube was heated to 1000~$^{\circ}$C at a rate of 100~$^{\circ}$C per hour, held for 6 hours at 1000~$^{\circ}$C, then the tube was cooled down to 900~$^{\circ}$C with a cooling rate of 1.25~$^{\circ}$C per hour.
Consequently, the tube was centrifuged at 900 $^{\circ}$C to separate the flux from the crystal. The hexagonal-shaped single crystals were obtained with dimensions of approximately 1$\times$1$\times$1 mm$^3$, as shown in the inset \cref{fig1}\textcolor{blue}{b}.
The phase purity of the crystal was confirmed by the X-ray diffraction (RIGAKU, Miniflex 600-C), Laue backscattering image (RIGAKU, RASCO-BLII) and energy dispersive X-ray spectroscopy (EDS) [Tungsten scanning electron microscope (W-SEM), Model: JSM-6010LA; JEOL].
The magnetisation measurements were carried out using a SQUID magnetometer (Quantum design, MPMS-XL), and electrical resistivity was measured using MPMS-XL with a custom-made transport probe \cite{paul2025growth,paul2025nonanalytic,paul2026multi,paul2026CrSb,wu2025fermi,wvs6hqfv}.
Temperature-dependent magnetization was measured using zero-field-cooled (ZFC) warming and field-cooled (FC) warming protocols at a temperature sweep rate of 1 K/min.

\section{ RESULTS AND DISCUSSION}
\subsection{Structural characterization}
Mn$_{3-x}$Co$_{x}$Sn ($x$ = 0, 0.25) crystallises in the hexagonal structure with space group P6$_3$/mmc  with typical dimensions of approximately 1$\times$1$\times$1 mm$^3$, as illustrated in the inset of \cref{fig1}\textcolor{blue}{b}.
The Mn and Co atoms occupy the 6h Wyckoff position, whereas the Sn atoms are located at the 2d site of the hexagonal unit cell.
The crystallographic orientation was determined using Laue diffraction method for both $x$ = 0 and $x$ = 0.25 crystals, as shown in \cref{fig1}\textcolor{blue}{a} and  \cref{fig1}\textcolor{blue}{b}.
 The stoichiometric composition is further verified by the EDS analysis, which revealed the sample composition as Mn$_{3.03}$Sn$_{0.97}$ and Mn$_{2.74}$Co$_{0.26}$Sn.
To confirm the phase purity, the grown single crystals were crushed into a fine powder and examined by powder X-ray diffraction.
To verify the phase purity and obtain the lattice parameter, Rietveld refinement of the diffraction pattern is carried out using the FullProf suite \cite{rodriguez1993recent,paul2025growth,paul2026CrSb} as shown in \cref{fig1}\textcolor{blue}{c} and \cref{fig1}\textcolor{blue}{d}.
The refined lattice parameters are tabulated in \cref{tab1}.
Upon Co doping, the lattice contracts by approximately 0.17\% and 0.22\% along the $a$ and $c$ axes, respectively.
Effectively, the unit cell volume contracts by 0.2\%, owing to the smaller atomic radius of Co compared to Mn.
In Mn$_3$Sn, DOS at the Fermi level ($E_\mathrm{F}$) exhibits a sharp peak \cite{han2025high}. DOS at the Fermi level can be reduced by pressure/strain, lowering the effective carrier density and increasing the resistivity \cite{han2025high}. First-principles calculations further validate such scenarios.
A similar effect is expected for Co doping, where the ~0.6\% bond-length contraction modifies interaction potential. It also enhances disorders.
The Mn–Mn–Mn bond angle decreases by 1.3\% with Co doping, which can modify the Mn–Mn exchange interactions and effectively contribute to the stabilization of the WFM state.
The effect of such lattice distortion and disorder can be probed through resistivity and magnetic measurements.
\begin{table}
    \caption{{Obtained Rietveld refined lattice parameters from x-ray diffraction comparing Mn$_3$Sn with Mn$_{0.25}$Co$_{0.25}$Sn}.}
    \label{tab1}
    \centering
    \begin{tabular}{cccc} 
 	\hline
    \hline
     Parameters & Mn$_3$Sn& Mn$_{0.25}$Co$_{0.25}$Sn  \\
     \hline
     $a=b$ ($\mathring{\text{A}}$) & 5.676(2) & 5.661(2)\\
     $c$ ($\mathring{\text{A}}$) & 4.531(2) & 4.521(2)\\
     $V$ ($\mathring{\text{A}}^3$) & 126.280(1) & 126.031(1)\\
    $c/a$ & 0.7982(1) & 0.7986(1)\\
    \ldelim\{{3}{*}[Bond angle at $z=0$ (degree)] & 17.88(3) & 17.64(4) \\
                      & 42.12(3) & 42.36(7)\\
                        & 120(0) & 120(0)\\
    Bond angle at $z=1/2$ & 60(0) & 60(0)\\
     Bond length at $z=0$ ($\mathring{\text{A}}$) &2.400(5)& 2.400(5)\\
    Bond length at $z=1/2$ ($\mathring{\text{A}}$) & 3.103(4)& 3.085(2)\\
     \hline
     \hline
     \end{tabular}
\end{table}
\begin{figure*}[t]
		\begin{center}
			\includegraphics[scale=0.67]{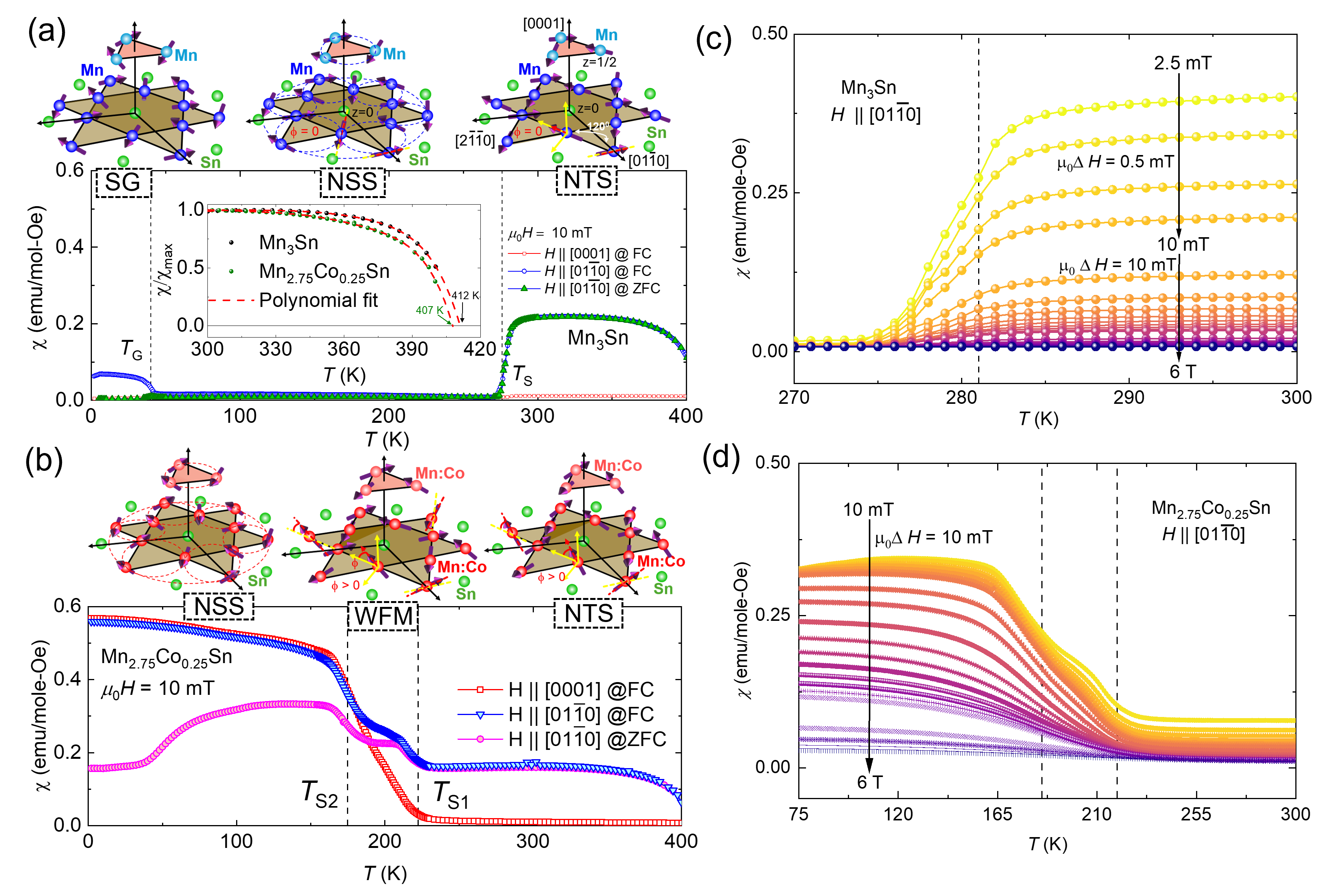}
			\caption{Temperature-dependent dc-susceptibility ($\chi$ = $M/H$) of (a) Mn$_3$Sn and (b) Mn$_{2.75}$Co$_{0.25}$Sn single crystals, measured under ZFC (blue) and FC (red) conditions in an applied field of 10 mT along the [01$\bar{1}$0] direction. Schematic diagrams of the possible magnetic spin configurations (corresponding to NTS, NSS, WFM, and SG phases) are also embedded with this $\chi-T$ curve. The dashed circle is made to display spin-spiral configuration. Upon Co doping, the magnetic moments tilt away from the kagome ($ab$) plane toward the $c$-axis by a canting angle $\phi~>~0$. The inset of (a) shows the estimation of $T_\mathrm{N}$ using a polynomial fit. (c-d) $\chi(T)$ dependence under various applied fields of 0.01–6 T across SRT.}
			\label{fig2}
		\end{center}
\end{figure*}
\subsection{Resistivity}
Figures \ref{fig1}\textcolor{blue}{e} and \ref{fig1}\textcolor{blue}{f} show the temperature-dependent zero-field longitudinal resistivity ($\rho_{zz}$) measured with current along the [0001] direction for Mn$_3$Sn and Mn$_{2.75}$Co$_{0.25}$Sn single crystals, respectively, in the temperature range 2–300 K.
Electrical contacts were made using silver epoxy (EPO-TEK, H20E), and the resistivity was measured using a standard linear four-probe technique \cite{paul2025growth,paul2025nonanalytic,paul2026multi,paul2026CrSb,wu2025fermi,wvs6hqfv}.
Mn$_3$Sn crystal exhibits \mbox{metallic} behaviour across the entire temperature range with a residual resistivity ratio ($\rho_\mathrm{300\  \rm{K}}/ \rho_{2\ \rm{K}}$) of 11.
The anomaly near $T_\mathrm{S}$ = 278 K corresponds to spin-reorientation transition (SRT) from low temperature NSS to high temperature NTS structure \cite{sung2018magnetic,nakatsuji2015large}.
The transition at 278 K further supports the stoichiometric composition of Mn$_3$Sn, as Mn-deficient Mn$_3$Sn exhibits comparatively lower $T_\mathrm{S}$ ($\sim$ 260 K) \cite{low2024topological, sung2018magnetic}.
A clear thermal hysteresis is evident upon heating and cooling, opening at 273 K and closing at 280 K, characteristic of FOPT \cite{sung2018magnetic,low2024topological}, as shown in the inset of \cref{fig1}\textcolor{blue}{e}.
Figure \ref{fig1}\textcolor{blue}{f} shows the temperature-dependent resistivity of Mn$_{2.75}$Co$_{0.25}$Sn during the heating and cooling cycle. 
The low-temperature resistivity can be well fitted using $\rho$($T$) = $\rho_0$ + $AT$ + $BT^2$, where $\rho_0$ is the residual resistivity, and the linear-in-$T$ term arising from electron–phonon scattering dominates over the quadratic contribution. Upon Co doping, the residual resistivity increases from 21 ${\mu} \Omega$ cm (Mn$_3$Sn) to 16 m$\Omega$ cm with RRR $\sim$ 1.8, attributed to the reduction of the DOS at $E_\mathrm{F}$ \cite{han2025high} owing to lattice contraction and enhanced disorder-induced scattering.
It is important to note that Kondo-like behaviour has been reported in Fe doped Mn$_3$Sn \cite{low2022tuning,low2024topological}; likely such behaviour is due to Fe induced magnetic inhomogeneity or cluster formation. 
The absence of Kondo-like behaviour in Mn$_{2.75}$Co$_{0.25}$Sn suggests a homogeneous nature of the grown crystal.
The EDS mapping further validates the homogeneous distribution of Mn, Co, and Sn throughout the crystal, without any Co clustering.
Unlike pristine Mn$_3$Sn, which exhibits a sharp resistivity jump at $T_\mathrm{S2}$, Mn$_{2.75}$Co$_{0.25}$Sn shows a continuous change in slope near 210 K, followed by a pronounced kink at 175 K, accompanied by substantial thermal hysteresis.
A clear hysteretic behaviour between 170 and 205 K as shown in the inset of \cref{fig1}\textcolor{blue}{f}, indicate signature of FOPT. 
Additionally, a change in slope around 210 K is likely driven by Co induced modifications of magnetic exchange that enhance weak ferromagnetism.

\begin{figure*}
		\begin{center}
			\includegraphics[scale=0.65]{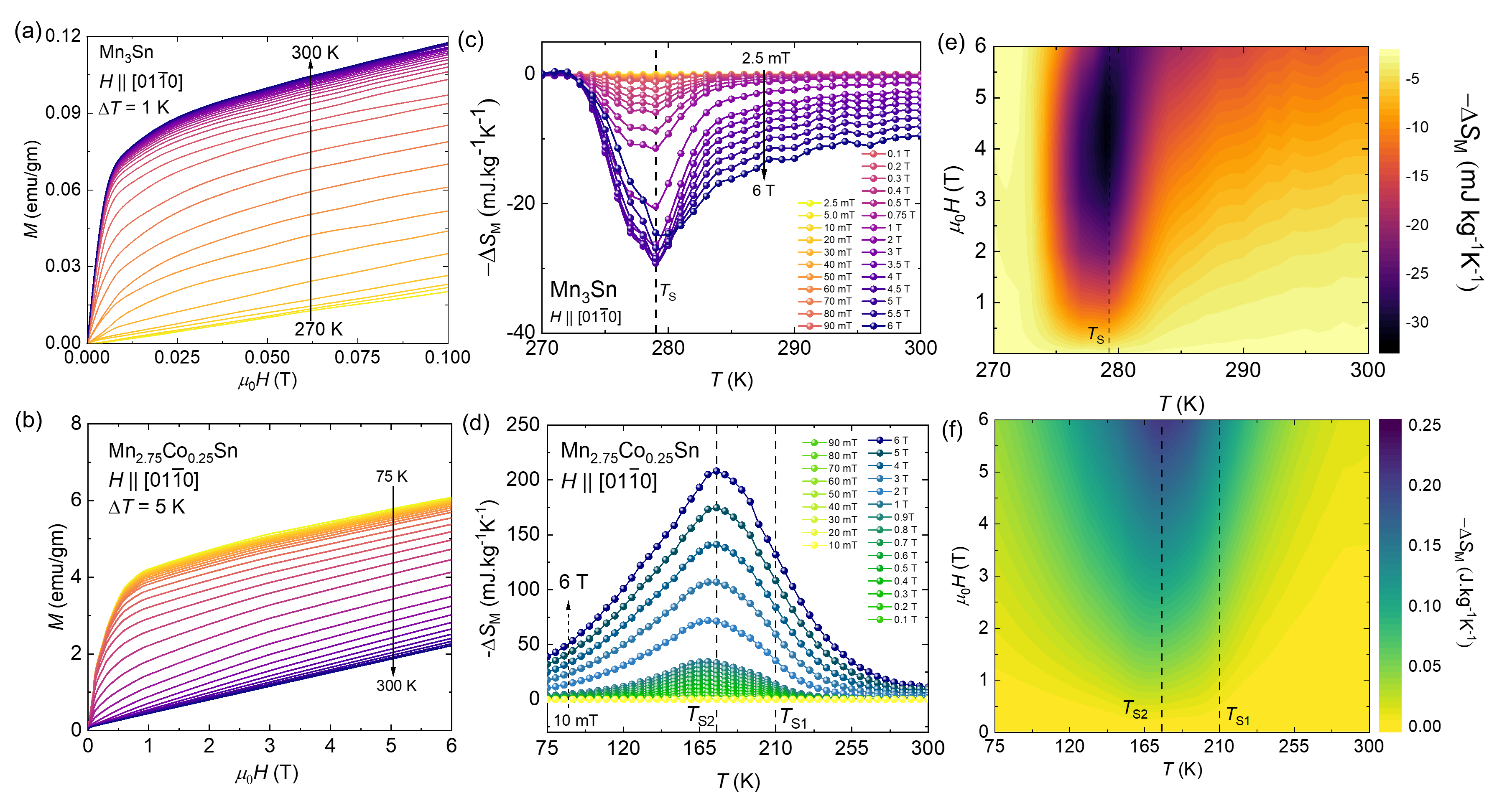}
			\caption{Isothermal magnetisation [$M(H)$] curves measured for (a) Mn$_3$Sn and (b) Mn$_{2.75}$Co$_{0.25}$Sn crystals. (c-d) Magnetic entropy change ($-\Delta S_\mathrm{M}$) as a function of temperature at various magnetic fields for both crystals. (e-f) 2D Colour contour maps of $-\Delta S_\mathrm{M}$ as a function of temperature and magnetic field for both crystals, highlighting the IMCE and CMCE, respectively.}
			\label{fig3}
		\end{center}
\end{figure*}
\subsection{Magnetism}
Mn$_3$Sn exhibits a strongly temperature-dependent magnetic phase with a non-collinear antiferromagnetic structure as shown in \cref{fig0}.
To investigate the impact in temperature-dependent magnetic behaviour due to Co doping on Mn$_3$Sn, ZFC and FC dc-susceptibility ($\chi = M/H$) is measured under an applied magnetic field of 100 Oe along \mbox{$H$ $\|$ [01$\bar{1}$0]} direction over the temperature range 1.8-400 K as shown in \cref{fig2}\textcolor{blue}{a} and \cref{fig2}\textcolor{blue}{b}.
The magnetic transition temperatures are determined from d$\chi$/d$T_\mathrm{{max}}$.
In Mn$_3$Sn, three magnetic phase transitions are observed at $T_\mathrm{G}$ = 40 K, $T_\mathrm{S}$ = 278 K and $T_\mathrm{N}$ $>$ 400 K as shown in \cref{fig2}\textcolor{blue}{a}.
The bifurcation of the ZFC and FC magnetisation curves at low temperatures suggests the presence of magnetocrystalline anisotropy. 
The transition at $T_\mathrm{G}$ = 40 K is attributed to disordered spin-glass (SG) behaviour, corresponding to the coexistence of AFM and long-range weak FM components \cite{sung2018magnetic,feng2006glassy,yan2019room}.
Upon heating, Mn moment undergoes changes from NSS to NTS structure above $T_\mathrm{S}$ = 278 K as schematically shown in \cref{fig2}\textcolor{blue}{a} (top).
The decrease in $\chi$ above 400 K indicates proximity to the Néel temperature, suggesting the transition from an antiferromagnetically ordered triangular arrangement of Mn moments to the paramagnetic state (\cref{fig2}\textcolor{blue}{a}) \cite{nakatsuji2015large,sung2018magnetic,yan2019room}.
Due to instrumental limitations, the measurement was carried out up to 400 K and $T_\mathrm{N}$ is estimated using polynomial fits, yielding 412 K as shown in the inset of \cref{fig2}\textcolor{blue}{a}).
Upon Co doping, $T_\mathrm{N}$ is effectively reduced (estimated to be $\sim$ 407 K).
Figure \ref{fig2}\textcolor{blue}{b} shows $\chi - T$ curve for Mn$_{2.75}$Co$_{0.25}$Sn single crystal.
A WFM transition ($T_\mathrm{S1}$) appears near 210 K, as shown in \cref{fig2}\textcolor{blue}{b}, attributed to spin canting.
Below $T_\mathrm{S1}$, the magnetisation increases with decreasing temperature, indicating that Mn$_{2.75}$Co$_{0.25}$Sn first undergoes a ferromagnetic-like transition followed by SRT. Such behaviour suggests the coexistence of dominant FM moments aligned along the $c$-axis, and AFM moments within the kagome plane, resulting in a net moment canted toward the $c$ direction as shown in \cref{fig2}\textcolor{blue}{b}.
In this regard, recent works on Fe-doped Mn$_3$Sn support such scenarios \cite{liu2020magnetic,hanggai2025complex,singh2024higher,tagami2024magnetic}.
The possible reason for the decrease in $T_\mathrm{S}$ upon Co doping is related to kagome lattice symmetry-breaking effect, which effectively destabilises DMI, nearest-neighbour Heisenberg exchange, and magnetic anisotropy energy.
Magnetic anisotropy is quantified with direction-dependent susceptibility ratio at temperature $T$, $\chi^\mathrm{R}_T~=~\chi_{[01\bar{1}0]}/\chi_{[0001]}$. For Mn$_3$Sn, $\chi^\mathrm{R}_\mathrm{300~K}$ is found to be 18, which is reduced to 15 upon Co doping.~At low temperature, pristine Mn$_3$Sn exhibits weak magnetic anisotropy with $\chi^\mathrm{R}_\mathrm{150~K}~=~1.8$.
In case of  Mn$_{2.75}$Co$_{0.25}$Sn, $\chi^\mathrm{R}_\mathrm{150~K}$ drastically reduces to 0.9, indicating a reorientation of the magnetic easy axis from the $ab$-plane toward the $c$-axis. A similar trend is observed for Mn$_{2.65}$Fe$_{0.25}$Sn \cite{low2024topological}, with $\chi^\mathrm{R}_{T}$ decreasing from 5.8 (300 K) to 0.4 (2 K).
Low-temperature susceptibility ($\chi_{2~\mathrm{K}}$)of Mn$_{2.75}$Fe$_{0.25}$Sn is three times higher than Mn$_{2.75}$Co$_{0.25}$Sn ($\sim$0.5 emu/mol-Oe) \cite{low2022tuning,low2024topological}.
Such high susceptibility with Kondo-like behaviour in Mn$_{2.75}$Fe$_{0.25}$Sn is likely due to the Fe clusters.
In addition, Co doping also suppresses the unwanted SG behaviour, possibly due to the reduction of intersite mixing of Mn with Sn.

The temperature dependence of the dc susceptibility was measured under ZFC conditions at fields of 2.5 mT and 6 T, as shown in \cref{fig2}\textcolor{blue}{c} and \cref{fig2}\textcolor{blue}{d} for Mn$_3$Sn and Mn$_{2.75}$Co$_{0.25}$Sn, respectively.
The decrease in susceptibility with increasing magnetic field arises from the saturation of the NSS structure and the suppression of spin fluctuations at high fields.
With increasing magnetic field in the Mn$_{2.75}$Co$_{0.25}$Sn, the spin reorientation and the WFM transition progressively merge into a single transition as shown in \cref{fig2}\textcolor{blue}{d} \cite{sarkar2008field,heritage2020images,tsukamoto2011magnetic}.
Now, the important question which arises: what is the nature of the phase transition under a high magnetic field?
So, the nature of the phase transition is studied through a powerful thermodynamic probe, namely MCE.
\begin{figure*}
		\begin{center}
			\includegraphics[scale=0.7]{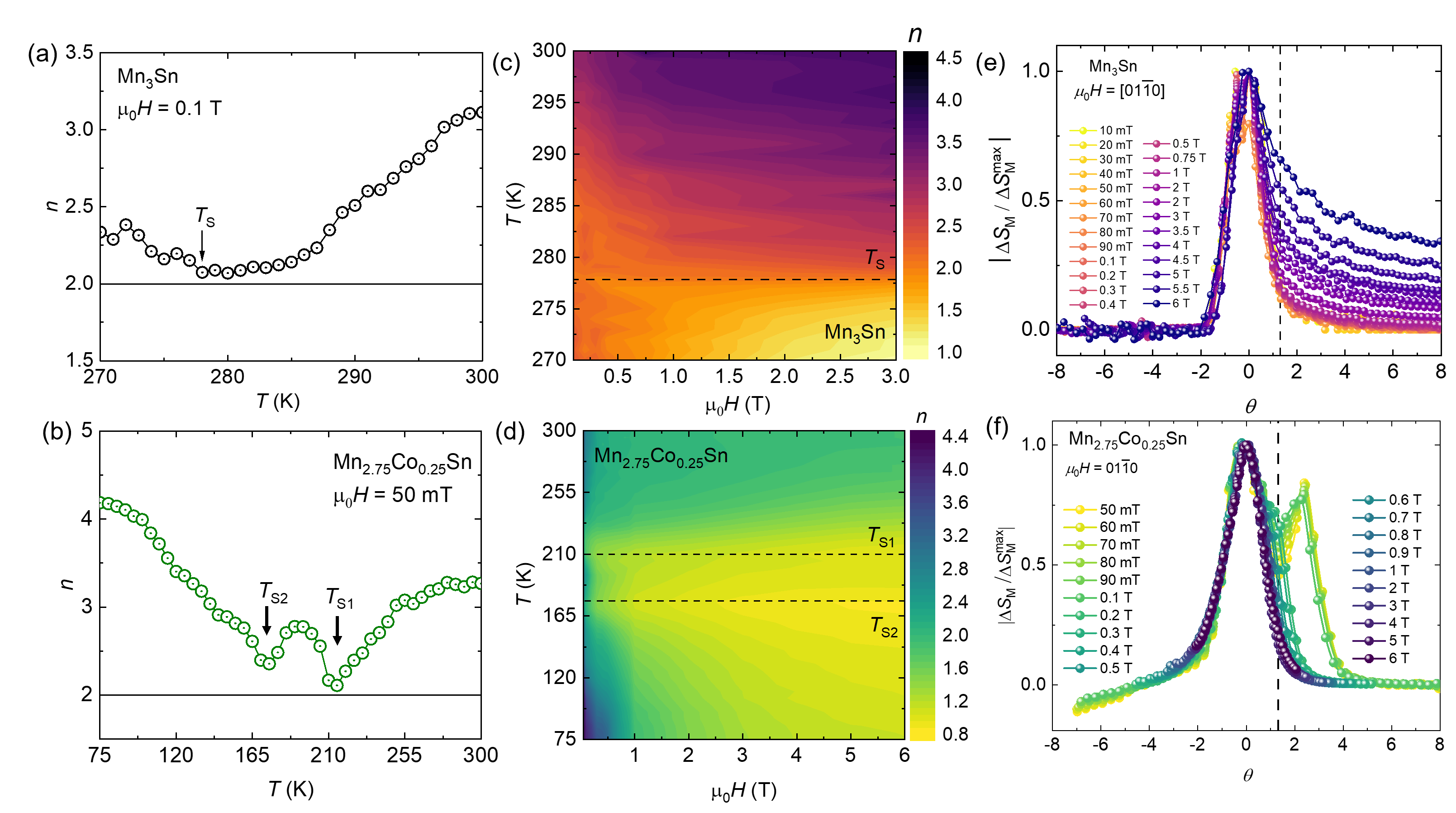}
			\caption{Field exponent $n$ of the magnetic entropy changes as a function of temperature at low field for (a) Mn$_3$Sn and (b) Mn$_{2.75}$Co$_{0.25}$Sn. Typically, $n > 2$ indicates FOPT type, whereas $0 < n < 2$ corresponds to SOPT.  (c-d) 2D color contour maps of field exponent $n(T,H)$ for both crystals. (e-f) Scaled magnetic entropy change, $\Delta S_\mathrm{M}/\Delta S_\mathrm{M}^\mathrm{max}$, plotted as a function of the rescaled temperature $\theta$. It is evident that for Mn$_3$Sn, the rescaled entropy curves at various fields do not collapse (for example, at $\theta$ = 1.3 as shown by a dotted line), suggesting FOPT.
            For Mn$_{2.75}$Co$_{0.25}$Sn, the rescaled entropy curves above 2 T collapse onto a universal curve, indicating a field-driven crossover from a FOPT to SOPT.}
			\label{fig4}
		\end{center}
\end{figure*}
\subsection{Magnetocaloric effect (MCE)}
The MCE refers to the magnetic entropy change ($-\Delta S_\mathrm{M}$) induced by applying or removing an isothermal magnetic field.
To study the MCE, isothermal magnetization ($M$–$H$) measurements were performed across SRT for Mn$_3$Sn and Mn$_{2.75}$Co$_{0.25}$Sn crystals along $H$ $\|$ [01$\bar{1}$0], as shown in \cref{fig3}\textcolor{blue}{a} and \cref{fig3}\textcolor{blue}{b}.
The $\Delta S_\mathrm{M}$ is evaluated from isothermal magnetization curves [$M(H)$] using the standard Maxwell thermodynamic relation \cite{gschneidner2005recent}.
\begin{equation}
    \Delta S_\mathrm{M} (T, H) = \mu_0 \int_{0}^{H} \left(\frac{\partial M\ (T, H)}{{\partial T}}\right)_H \ \mathrm{d}H
    \label{eq1}
\end{equation}
Since the isothermal $M$–$H$ data are obtained at discrete field intervals, the integral can be approximated using the following numerical expression \cite{sahoo2018influence,tegus2002magnetic}:
\begin{equation}
    -\Delta S_\mathrm{M} (T,H) = \sum_{m}\frac{M_{m} \ (T_{m},H)-M_{m+1}\ (T_{m+1}, H)}{T_{m}-T_{m+1}}\Delta H_{m}
    \label{eq2}
\end{equation}
where $M_{m}$ and $M_{m+1}$ denote the magnetization measured at temperatures $T_{m}$ and $T_{m+1}$, respectively, and $\Delta H_{m}$ corresponds to the field variation between $H = 0$ and $H_\mathrm{max}$.
The value of $-\Delta S_\mathrm{M}$ at the average temperature $T_\mathrm{ave} = (T_{m} + T_{m+1})/2$ is obtained from the total area enclosed between two successive isothermal $M$–$H$ curves measured at $T_{m}$ and $T_{m+1}$.
Figures \ref{fig3}\textcolor{blue}{c} and \ref{fig3}\textcolor{blue}{d} show the temperature dependence of $-\Delta S_\mathrm{M}$ at various fields for Mn$_\mathrm{3}$Sn and Mn$_\mathrm{2.75}$Co$_\mathrm{0.25}$Sn crystals, respectively, calculated using \cref{eq2}.
The negative value of $-\Delta S_\mathrm{M}$ in Mn$_3$Sn across $T_\mathrm{S}$ is attributed to IMCE associated with NSS to NTS structure.
Mn$_3$Sn exhibits a maximum magnetic entropy change at 278 K with ($|\Delta S_\mathrm{M}|$) of $\approx$ 29 mJ kg$^{-1}$ K$^{-1}$ up to 4 T but gradually decreases with increasing field.
The reduction of $|\Delta S_\mathrm{M}|$ above 4 T is attributed to the evolution of magnetic domain walls, where local variations in magnetisation lead to a smooth spin reorientation along the field direction \cite{yan2019room,sung2018magnetic,stoner1948mechanism}.
With 8\% Co doping, Mn$_3$Sn undergoes a transition from IMCE to CMCE, with a maximum entropy change ($|\Delta S_\mathrm{M}^\mathrm{max}|$) of about 200 mJ kg$^{-1}$ K$^{-1}$, as shown in \cref{fig3}\textcolor{blue}{d}.
Although, Mn$_{3-x}$Fe$_x$Sn ($x = 0.8–1.4$) exhibits a $|\Delta S_\mathrm{M}^\mathrm{max}|$ of 1100–1410 mJ kg$^{-1}$ K$^{-1}$, increasing by about 30\% with Fe concentration \cite{hanggai2025complex}. Relatively smaller value (nearly six times) of $|\Delta S_\mathrm{M}^\mathrm{max}|$ in Mn$_{3-x}$Co$_{x}$Sn ($x=0.25$) is primarily attributed to the concentration effect rather than an intrinsic difference between Co and Fe doping.
Figures \ref{fig3}\textcolor{blue}{e} and \ref{fig3}\textcolor{blue}{f} display the $H-T$ phase diagrams of $-\Delta S_\mathrm{M}$ for Mn$_\mathrm{3}$Sn and Mn$_\mathrm{2.75}$Co$_\mathrm{0.25}$Sn crystals, respectively.
Mn$_3$Sn shows a sharp peak in $-\Delta S_\mathrm{M}$, nearly FWHM within 5 K, typical of a FOPT.
In contrast, Mn$_\mathrm{2.75}$Co$_\mathrm{0.25}$Sn exhibits a broader dip (FWHM = 94 K) corresponding to two magnetic transitions ($T_\mathrm{S2}$ = 175 K, $T_\mathrm{S1}$ = 210 K) as shown in \cref{fig3}\textcolor{blue}{f}.

The magnetic entropy change under an applied magnetic field adheres to a power-law: $|\Delta S_\mathrm{M}|$ $\propto$ $H^{n}$.
The field exponent $n$ serves as an indicator of the order of the magnetic phase transition.
The temperature and field dependence of $-\Delta S_\mathrm{M}$ differ significantly between IMCE and CMCE materials.
To probe the nature of the phase transition locally at each temperature and field, $n(T,H)$ is analyzed using the relation:
\begin{equation}
n(T,H) = \frac{\mathrm{d}(\ln~|\Delta S_\mathrm{M}|)}{\mathrm{d}(\ln~H)}.
\end{equation}
\noindent The value of $n(T,H)$ lies between 0 and 2 for SOPT, whereas $n>2$ is characteristic of FOPT \cite{romero2017two,law2018quantitative,hanggai2025complex}.
Figures \ref{fig4}\textcolor{blue}{a} and \ref{fig4}\textcolor{blue}{b} show the temperature dependence of $n$ at low  field for Mn$_3$Sn and Mn$_{2.75}$Co$_{0.25}$Sn, respectively.
For Mn$_3$Sn, $n(T,\mu_0H=0.1~\mathrm{mT})>2$ throughout the temperature evolution with a sharper change near $T_\mathrm{S}$ [\cref{fig4}\textcolor{blue}{a}], indicating a signature of FOPT \cite{sung2018magnetic, law2018quantitative,bonilla2010universal}.
In case of Mn$_{2.75}$Co$_{0.25}$Sn, $n(T,\mu_0H=50~\mathrm{mT})$ displays two minima at 175 and 210 K [\cref{fig4}\textcolor{blue}{b}], with $n>2$, suggesting FOPTs.
Further, to demonstrate the temperature and field evolution of $n(T,H)$, the $T-H$ phase diagrams for Mn$_3$Sn and Mn$_{2.75}$Co$_{0.25}$Sn are displayed in Figs. \ref{fig4}\textcolor{blue}{c} and \ref{fig4}\textcolor{blue}{d}, respectively.
Due to the complex nature of the IMCE for Mn$_3$Sn at high fields, we evaluate $n(T,H)$ up to 3 T. The value of $n(T_\mathrm{S},H)$ remains constant, indicating the absence of a field-induced phase transition crossover in Mn$_3$Sn [\cref{fig4}\textcolor{blue}{c}]. 
Interestingly, for Mn$_{2.75}$Co$_{0.25}$Sn, the field-exponent undergoes a change from $n(T,H)>2$ to $n(T,H)<2$, depicting a field-driven crossover from FOPT to SOPT~[\cref{fig4}\textcolor{blue}{d}].
The characteristics of such dual ordering are further elucidated by extracting the critical exponents across SRT. Relative cooling power (RCP) is defined as RCP = $-\Delta S_\mathrm{M}^\mathrm{max} \times \Delta T_\mathrm{FWHM}$ ($\Delta T_\mathrm{FWHM}$ = full width at half maximum of the $-\Delta S_\mathrm{M}$ vs $T$ curve), follows a power-law dependence on the magnetic field; $\mathrm{RCP} \propto H^{1+1/\delta}$, where $\delta$ represents the critical exponent of the magnetic transition with $\delta = 1+ \gamma/\beta$.  The exponent $n$ is connected to the critical exponents by $n = 1 + (\beta - 1)/(\beta + \gamma)$. The power-law fitting of the field-dependent RCP and $|\Delta S_\mathrm{M}|$ yields $\delta$ = 4.89 and $n$ = 0.86, corresponding to $\beta$ = 0.6 and $\gamma = 2.3$. 
The Widom scaling law yields $\delta=4.83$, consistent with the $\delta$ value obtained from the RCP. 
The deviation of the obtained critical exponents from known universality classes is possibly due to the coexistence of FOPT and SOPT and the presence of competing WFM and AFM interactions induced by Co doping in Mn$_3$Sn \cite{mohan1998critical,hanggai2025complex,gaur2023observation}.

A phenomenological universal curve is constructed by scaling the magnetic entropy change ($\Delta S_\mathrm{M}/\Delta S^\mathrm{max}_\mathrm{M}$), such that data corresponding to equivalent thermodynamic states collapse onto a single curve in the case of SOPT when plotted as a function of the reduced temperature ($\theta$), defined as \cite{bonilla2010universal,ali2019rotating,franco2010scaling}
\begin{equation}
\theta =
\begin{cases}
-\dfrac{T - T_\mathrm{peak}}{T_{r_1} - T_\mathrm{peak}}, & T \leq T_\mathrm{peak} \\
\\
\dfrac{T - T_\mathrm{peak}}{T_{r_2} - T_\mathrm{peak}}, & T > T_\mathrm{peak}.
\end{cases}
\label{eq3}
\end{equation}
\noindent Here, $T_{r_1}$ and $T_{r_2}$ correspond to the temperatures at the full width at half maximum of the $\Delta S_\mathrm{M}$ anomaly ($\Delta S^\mathrm{max}_\mathrm{M}/2$).
Figures \ref{fig4}\textcolor{blue}{e} and \ref{fig4}\textcolor{blue}{f} show the $|\Delta S_\mathrm{M}/\Delta S^\mathrm{max}_\mathrm{M}|$ versus $\theta$ at different magnetic fields for Mn$_3$Sn and Mn$_{2.75}$Co$_{0.25}$Sn, respectively.
For Mn$_3$Sn, $\Delta S_\mathrm{M}/\Delta S^\mathrm{max}_\mathrm{M}$ curves for various magnetic fields deviate from the universal curve, signifying FOPT \cite{franco2010scaling,franco2006field, krenke2005inverse,biswas2013universality,bonilla2010universal,law2018quantitative,nakatsuji2015large,sung2018magnetic}.
For Mn$_{2.75}$Co$_{0.25}$Sn, two distinct peaks are clearly observed in the scaled magnetic entropy change, corresponding to two successive magnetic phase transitions, as shown in \cref{fig4}\textcolor{blue}{f}.
A field-dependent collapse of $|\Delta S_\mathrm{M}/\Delta S^\mathrm{max}_\mathrm{M}|$ is noticed.
In particular, field-dependent scaled entropy curves collapse onto a single universal curve above 2T, suggesting the field-driven crossover from FOPT to SOPT. However, there is a noticeable deviation below 2~T, indicating the coexistence of SOPT and FOPT, as illustrated in  \cref{fig4}\textcolor{blue}{f}.
In order to quantify the uncertainty associated with the collapse of the scaled $\Delta S_\mathrm{M}(T)$ curves, a dispersion parameter for each point of the universal curve is defined as \cite{bonilla2010universal,ali2019rotating,franco2010scaling}:
\begin{equation}
d = 100 \times \frac{W}{(\Delta S_\mathrm{M}/\Delta S^\mathrm{max}_\mathrm{M})},
\label{eq4}
\end{equation}
where $W$ represents the vertical deviation of each scaled entropy curve from the mean universal curve.
In materials undergoing FOPT, the dispersion often exceeds 100\%, whereas SOPT exhibits values below 30\%, which are generally attributed to measurement uncertainty.
For Mn$_3$Sn, the dispersion ($d$) is 104\%, indicating that the IMCE corresponds to a FOPT. In Mn$_{2.75}$Co$_{0.25}$Sn, the dispersion reduces from 73\% to below 5\% with increasing the fields, suggesting a field-driven crossover from FOPT to SOPT.
\begin{figure}
	\begin{center}
			\includegraphics[scale=0.5]{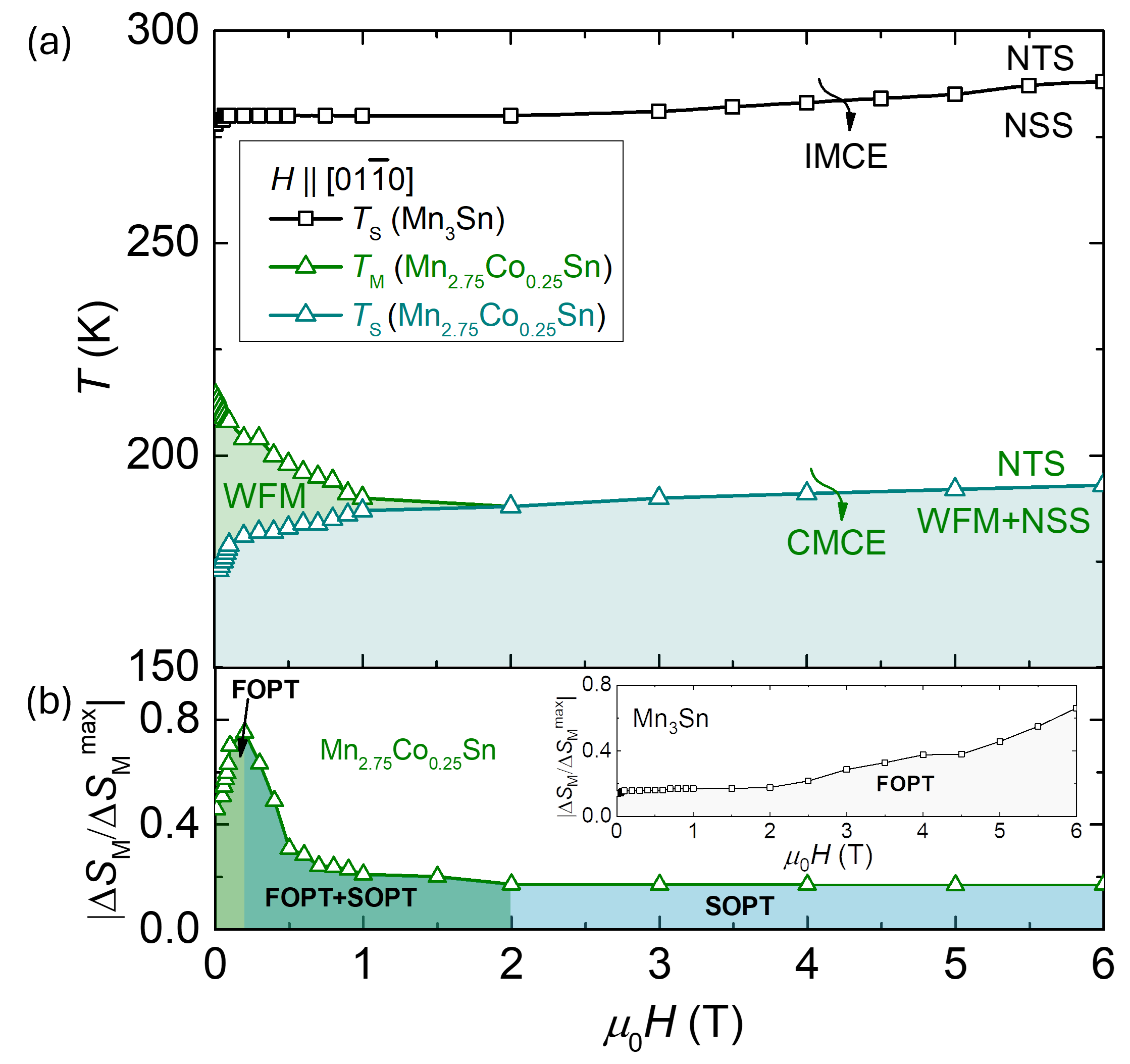}
			\caption{Field-dependent (a) magnetic and (b) MCE phase diagrams of Mn$_3$Sn and Mn$_{2.75}$Co$_{0.25}$Sn single crystal with $H\ \|\ [01\bar{1}0]$. The values of $|\Delta S_\mathrm{M}/\Delta S^\mathrm{max}_\mathrm{M}|$ for different fields obtained at $\theta\ = 1.3$ as indicated by dashed line in \cref{fig4}\textcolor{blue}{e} and \cref{fig4}\textcolor{blue}{f}.}
			\label{fig5}
		\end{center}
\end{figure}

Based on the magnetic measurements and MCE analysis, we propose magnetic and MCE phase diagrams of Mn$_{3-x}$Co$_x$Sn as shown in \cref{fig5}.
The field dependence of magnetic ordering temperatures (\cref{fig5}\textcolor{blue}{a}) is evaluated from $d\chi/dT_\mathrm{max}$.
For Mn$_3$Sn, a gradual increase in  $T_\mathrm{S}$ is noticed (from 278 K to 288 K) with a field change of 6 T.
Effect of Co doping reduces the  $T_\mathrm{S2}$ (175 K) and introduces WFM below $T_\mathrm{S1}$ = 210 K.
Interestingly, $T_\mathrm{S1}$ decreases while $T_\mathrm{S2}$ increases with field, and eventually merge with each other near 2 T. Above 2 T, it remains nearly constant.
Further, the field dependence of $|\Delta S_\mathrm{M}/\Delta S^\mathrm{max}_\mathrm{M}|$ at $\theta=1.3$ is plotted in \cref{fig5}\textcolor{blue}{b} to illustrate the phase transition crossover scenario. 
For Mn$_3$Sn, $|\Delta S_\mathrm{M}/\Delta S^\mathrm{max}_\mathrm{M}|$ increases monotonically with field, indicating FOPT as shown in the inset of \cref{fig5}\textcolor{blue}{b}.
For Mn$_{2.75}$Co$_{0.25}$Sn, $|\Delta S_\mathrm{M}/\Delta S^\mathrm{max}_\mathrm{M}|$ increases sharply up to 0.2 T (suggesting FOPT: highlighted in green) and decreases thereafter until 2 T,  attributed to the coexistence of FOPT and SOPT (highlighted in sea green).
Further increasing the field, all curves collapse onto a single universal master curve with nearly constant $|\Delta S_\mathrm{M}/\Delta S^\mathrm{max}_\mathrm{M}|$, signifying a field-driven crossover from FOPT to SOPT (highlighted in cyan).

\section{Conclusions}
In summary, we have successfully grown Mn$_{3-x}$Co$_x$Sn ($x$ = 0 and 0.25) single crystals and investigated their structure, magnetism and magnetocaloric properties. 
Pristine Mn$_3$Sn exhibits metallic behaviour with $T_\mathrm{S}$ = 278 K, whereas Co doping shifts $T_\mathrm{S}$ to 175 K, due to lattice contraction and enhanced disorder-induced scattering. Pronounced hysteresis in resistivity across the SRT indicates the characteristic of FOPT.
Co doping breaks the kagome lattice symmetry and induces an additional WFM transition at $T_\mathrm{S1} \approx 210$ K.
Such a complex nature of the magnetic phase transition is further studied through MCE.
A quantitative framework is developed to determine the order of \mbox{thermodynamic} phase transitions from the field dependence of MCE.
The magnetic entropy change ($-\Delta S_\mathrm{M}$) reveals a transition from IMCE to CMCE due to Co doping.
Most strikingly, a field-driven FOPT to SOPT is observed in Mn$_{2.75}$Co$_{0.25}$Sn.
The MCE phase diagram portrays FOPT, mixed and SOPT regions.

\subsection*{ ACKNOWLEDGMENTS}
We are grateful to Shingo Yonezawa, Yoshiteru Maeno, Giordano Mattoni, Atsutoshi Ikeda, Dipak Sahu, Bharti, Amitava Ghosh, and Wakeel Ahmad for useful discussions, helping with the Laue diffraction and magnetic measurements. C.S. acknowledges research support from IIT Kanpur Initiation Grant (IITK-2019-037) and research grants from Science and Engineering Research Board (SERB), Government of India (Grants No. CRG-2022-005726, EEQ-2022-000883).

\subsection*{DATA AVAILABILITY}
The data that support the findings of this study are available from the corresponding authors upon reasonable request.

\bibliography{ref}
\end{document}